\documentclass[aps, twocolumn, superscriptaddress, longbibliography]{revtex4-1}
\usepackage[english]{babel}
\usepackage{graphicx}
\usepackage{stmaryrd}
\usepackage{amsmath, amsfonts, amssymb}
\usepackage{mathrsfs}
\usepackage{braket}
\usepackage{color}
\usepackage{xspace}
\usepackage{amscd}
\usepackage{latexsym}
\usepackage{diagbox}
\usepackage{bm}
\definecolor{lightblue}{rgb}{0.13, 0.26, 0.99}

\usepackage[
colorlinks=true,
urlcolor=blue,
citecolor=blue,
linkcolor=blue,
hyperfootnotes=false]{hyperref}

\allowdisplaybreaks

\begin{document}
	\title{Universality of semisuper-Efimov effect}
	
	\author{Yusuke Horinouchi}
	\affiliation{RIKEN Center for Emergent Matter Science (CEMS), Wako, Saitama 351-0198, Japan}
	
	\date{\today}
	\begin{abstract} 
		We study the semisuper-Efimov effect, which is found for four identical bosons with a resonant three-body interaction in 2D, in various systems. Based on solutions of bound-state and renormalization-group equations, we first demonstrate an emergence of the semisuper-Efimov effect in mass-imbalanced bosons in 2D. Compared with the Efimov and the super-Efimov effects, the mass ratio-dependent scaling parameter is unexpectedly found to take on a finite value even for extremely mass-imbalanced situations, where the mass ratio is 0 or $\infty$. By a renormalization-group analysis, we also show that a weak two-body interaction sustains the semisuper-Efimov effect. Finally, we liberate the universality of the semisuper-Efimov effect from 2D by showing that bosons with linear-dispersion relation support the semisuper-Efimov effect in 1D. 
	\end{abstract}
	\maketitle
	\section{introduction}
	Efimov effect is one of prominent examples of universal phenomena that appear in resonantly interacting few-body systems. In the seminal paper by Efimov~\cite{efimov1970}, an emergence of an infinite series of self-similar trimer states is predicted for three-body systems with resonant $s$-wave interaction, irrespective of the specific form of the two-body interaction. Reflecting its self-similar binding energies $E_{n+1}/E_{n}\simeq (22.7)^{-2}$, the universality of the Efimov effect is represented by the renormalization-group limit cycle~\cite{bedaque1999, bedaque1999b, moroz2009, horinouchi2015}, which refers to a periodic renormalization-group flow~\cite{wilson1971, glazek2002}. Due to its universality and self similarity, the Efimov effect is extensively studied in variety of physical systems including nucleons~\cite{hammer2010}, mass-imbalanced fermions~\cite{efimov1973, petrov2003}, particles in mixed dimensions~\cite{nishida2009}, magnons~\cite{nishida2013b}, and macromolecules~\cite{pal2013}. In particular, experimental observations of the Efimov effect in ultracold atoms~\cite{kraemer2006, knoop2009, zaccanti2009, gross2009, pollack2009, huckans2009, barontini2009, pires2014, huang2014, tung2014, kunitski2015} provide a renewed interest to this subject 40 years after the Efimov's prediction. Among the theoretical studies in ultracold atoms is a prediction of super-Efimov effect \cite{nishida2013} in a two-dimensional system of three-spinless fermions with resonant $p$-wave interaction, where the three-body binding energy $E_n$ exhibits a double exponential growth $E_n\propto \exp(e^{\gamma n})$. The super-Efimov effect is shown to emerge also for mass-imbalanced systems in 2D provided that the two-body interaction is dominated by the $p$-wave contribution \cite{moroz2014}. \par
	In this paper, we focus on yet another few-body clusters proposed by Nishida \cite{nishida2017}. Motivated by theoretical proposals \cite{petrov2014, daley2014, petrov2014b, paul2016} for realizing a three-body interaction in absence of a two-body interaction, Nishida investigates four identical bosons with a resonant three-body interaction in 2D. He finds that the system supports the semisuper-Efimov tetramers where the four-body binding energy $E_n$ grows as $E_n\propto e^{(\pi n)^2}$. 
	Despite its qualitatively distinct feature from the Efimov and the super-Efimov effects, the semisuper-Efimov effect in other systems than identical bosons in 2D is yet to be explored. Therefore, we here extend the universality to mass-imbalanced systems and, in particular, demonstrate a qualitative difference of the mass-ratio dependent scaling parameter from those of the Efimov and the super-Efimov effects: The scaling parameter of the semisuper-Efimov effect is unexpectedly found to be stable against a variation of the mass ratio and takes on a finite value even in extremely mass-imbalanced situations where the mass-ratio takes on 0 or $\infty$. 
	To further demonstrate universality of the semisuper-Efimov effect, we also liberate the semisuper-Efimov effect from two-dimensional systems by showing an emergence of the semisuper-Efimov effect for bosonic particles with a linear-dispersion relation in 1D. 
	
	We organize the paper in the following manner: In Sec.~\ref{sec:mass-imbalanced}, we study mass-imbalanced bosons in 2D. Firstly, we investigate two-component mass-imbalanced bosons in 2D, as the simplest extension. Analytical solutions of bound-state and renormalization-group equations show an emergence of the semisuper-Efimov effect with a mass-ratio dependent scaling parameter. Effects of two-body interactions are then evaluated quantitatively by a renormalization-group analysis. As another extension, we also investigate three-component mass-imbalanced bosons. In Sec.~\ref{sec:1dlinear}, we further extend the universality to 1D by investigating identical bosons with a linear-dispersion relation. Similarly to Sec.~\ref{sec:mass-imbalanced}, we first show an emergence of the semisuper-Efimov effect in 1D by solving a bound-state equation and then discuss its stability against a weak two-body interaction by a renormalization-group analysis. Sec.~\ref{sec:summary} is devoted to the summary and discussion of this paper.
	
	\section{mass-imbalanced bosons in 2D}\label{sec:mass-imbalanced}
	\subsection{Two-component bosons}
	We first consider a system of two-component bosons in 2D with a resonant three-body interaction. With a model Hamiltonian, we first solve a four-body bound-state equation analytically and show that the system supports the semisuper-Efimov effect with a mass ratio-dependent scaling parameter. 
	We then support the analytical result and discuss its stability against two-body interactions by a renormalization-group analysis. 
	\subsubsection{Model analysis}\label{sec:2cbma}
	As a system of two-component bosons, we consider the following model Hamiltonian:
	\begin{widetext}
	\begin{align}
		H=\int\frac{d^2q}{(2\pi)^2}\frac{{\boldsymbol q}^2}{\alpha}\psi^{\dag}_{A,{\boldsymbol q}}\psi_{A,{\boldsymbol q}}
		+\int\frac{d^2q}{(2\pi)^2}{\boldsymbol q}^2\psi^{\dag}_{B,{\boldsymbol q}}\psi_{B,{\boldsymbol q}}
		-\frac{\lambda}{4}\int\frac{d^2Q d^2q'_x d^2q'_y d^2q_x d^2q_y}{(2\pi)^{10}}
		\chi({\boldsymbol q}'_x,{\boldsymbol q}'_y)\chi({\boldsymbol q}_x,{\boldsymbol q}_y)\nonumber\\
		\times
		\psi^{\dag}_{A, \frac{ \alpha {\boldsymbol Q}}{2\alpha+1}+{\boldsymbol q}'_y}
		\psi^{\dag}_{A, \frac{ \alpha {\boldsymbol Q}}{2\alpha+1}-\frac{\alpha{\boldsymbol q}'_y}{\alpha+1}+{\boldsymbol q}'_x}
		\psi^{\dag}_{B,\frac{\boldsymbol Q}{2\alpha+1}-\frac{{\boldsymbol q}'_y}{\alpha+1}-{\boldsymbol q}'_x}
		\psi_{B,\frac{\boldsymbol Q}{2\alpha+1}-\frac{{\boldsymbol q}_y}{\alpha+1}-{\boldsymbol q}_x}
		\psi_{A, \frac{ \alpha {\boldsymbol Q}}{2\alpha+1}-\frac{\alpha{\boldsymbol q}_y}{\alpha+1}+{\boldsymbol q}_x}
		\psi_{A, \frac{ \alpha {\boldsymbol Q}}{2\alpha+1}+{\boldsymbol q}_y},
		\label{eq:H_2bosons}
	\end{align}
	\end{widetext}
	where $\psi^{\dag}_A$ and $\psi^{\dag}_B$ ($\psi_A$ and $\psi_B$) represent the creation (annihilation) operators of the the two species $A$ and $B$ of bosons, respectively. The parameter $\alpha=\frac{m_{A}}{m_{B}}\in (0,\infty)$ is the mass-ratio between the two species of bosons. Here we employ the units $\hbar=2m_{B}=1$. We note that the Jacobi coordinate is employed in the three-body-interaction term since it facilitates loop-momentum integrals that appear in calculating $T$ matrices. The function $\chi$ represents a separable potential which is tractable due to its separability, and we here choose the following Lorentzian function:
	 \begin{align}
	 	\chi({\boldsymbol q}_x,{\boldsymbol q}_y)
		=\frac{\Lambda^2}{\frac{\alpha+1}{\alpha}q_x^2+\frac{2\alpha+1}{\alpha(\alpha+1)}q_y^2+\Lambda^2}.\label{eq:regulator}
	\end{align}
	To ensure the system to be Galiean invariant, we choose $\chi$ depending only on the relative momenta. In the limit of taking $\Lambda\rightarrow\infty$ ($\chi\rightarrow 1$), the three-body interaction in Eq.~(\ref{eq:H_2bosons}) reduces to the contact interaction. To put it another way, the separable potential $\chi$ plays the role of the short-range regulator of the system with the ultraviolet cutoff $\Lambda$. \par 
	 \begin{figure}[t]
  		\begin{center}
  			 	\includegraphics[width=240pt]{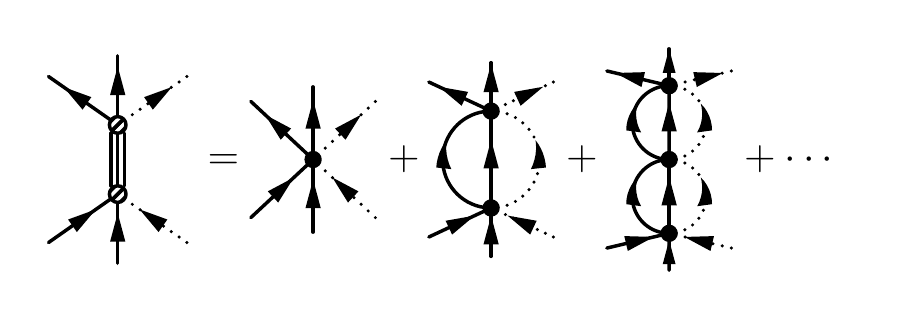}
				\caption{Diagrammatic expression of the three-body $T$-matrix $T^{(3)}$ for two-component bosons, where $T^{(3)}$ is represented by the triple-lines on the left-hand side. The shaded circle on the left-hand side represents the separable potential $\chi$ which is not renormalized by quantum fluctuations. Here solid and dashed lines represent propagators of the two species $A$ and $B$ of bosons, respectively.}\label{fig:fig1}
  			\end{center}
	\end{figure}
	\begin{figure}[t]
  		\begin{center}
  			 	\includegraphics[width=200pt]{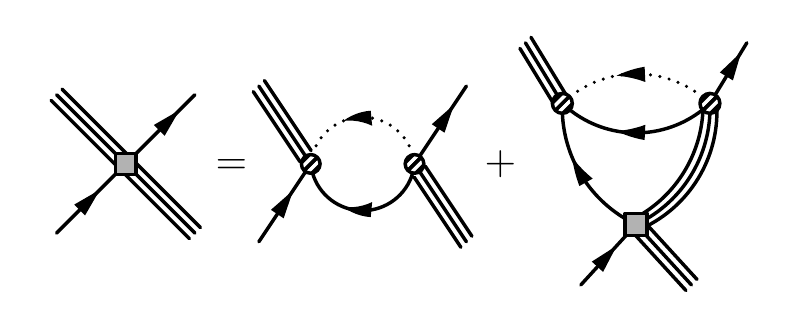}
				\caption{Skornyakov-Ter-Martirosian-type linear integral equation of the four-body $T$-matrix $T^{(4)}$ for the two-component bosons, where $T^{(4)}$ is represented by the shaded square on the left-hand side. The equation can be obtained by summing up the ladder-type Feynman diagrams that refer to the two-particle exchange processes.}\label{fig:fig2}
  			\end{center}
	\end{figure}
	Based on the model, we calculate the three- and the four-body $T$ matrices analytically by ressumming the ladder-type Feynman diagrams which are the only non-vanishing diagrams in the particle vacuum. Concerning the three-body sector, an exact $T$-matrix $T^{(3)}$ is given by the diagrams depicted in Fig.~\ref{fig:fig1}, and we thus obtain $T^{(3)}$ as 
	\begin{align}
		&T^{(3)}(p^0,{\boldsymbol p}; {\boldsymbol q}'_x,{\boldsymbol q}'_y, {\boldsymbol q}_x,{\boldsymbol q}_y)
		=\frac{\chi({\boldsymbol q}'_x,{\boldsymbol q}'_y)\chi({\boldsymbol q}_x,{\boldsymbol q}_y)}
			{-\frac{1}{g}-\frac{1}{32\pi^2}\frac{\alpha^2\epsilon}{2\alpha+1}f\left(\frac{\epsilon}{\Lambda^2}\right)},\label{eq:T_3}
		\\
		&f\left(x\right)=\frac{1}{1-x}\left(\frac{1}{1-x}\ln x+1\right)\underset{x\rightarrow0}{\sim}\ln x,\label{eq:funcf}
	\end{align}
	where $\epsilon=ip^0+\frac{p^2}{2\alpha+1}$ is the total energy of the three particles, and $g$ is the renormalized three-body coupling constant which is related to the bare coupling $\lambda$ via
	\begin{align}
		\frac{1}{\lambda}-\frac{1}{32\pi^2}\frac{\alpha^2}{2\alpha+1}\Lambda^2=\frac{1}{g}.
	\end{align} 
	A three-body bound state can be obtained as a pole of $T^{(3)}$ in Eq.~\eqref{eq:T_3}, and for a large-positive $g$, we find a shallow bound state whose energy $\epsilon$ satisfies $\epsilon \ln \frac{\epsilon}{\Lambda^2}=-\frac{32\pi^2(2\alpha+1)}{\alpha^2}\frac{1}{g}$. The resonance condition is, therefore, achieved by tuning $1/g=0$ so that the three-body binding energy $\epsilon$ vanishes. As a complement, we calculate the $T$ matrix $T^{(3)}$ with different separable potentials $\chi$ from Eq.~(\ref{eq:regulator}) and check that the asymptotic functional form of $f\left(\frac{\epsilon}{\Lambda^2}\right)\sim \ln\frac{\epsilon}{\Lambda^2}$ in Eq.~(\ref{eq:T_3}) does not depend on the specific choice of $\chi$ at sufficiently low-energy $\epsilon/\Lambda^2 \ll 1$. \par
	We then turn to the four-body sector where we consider the scattering of three identical bosons $\psi_A$ with a distinguishable boson $\psi_B$. In the system, the four-body $T$-matrix $T^{(4)}$ satisfies the Skornyakov-Ter-Martirosian-type \cite{skornyakov1956} integral equation depicted in Fig.~\ref{fig:fig2}. According to the Kallan-Lehmann spectral representation, four-body bound states are obtained from poles of $T^{(4)}$ with respect to the total energy $E$ of the four particles. 
	In particular, when the total energy $E$ approaches to one of four-body binding energies $E\rightarrow E_n$, $T^{(4)}$ factorizes as $T^{(4)}(E;{\boldsymbol q}',{\boldsymbol q})=\frac{Z({\boldsymbol q}')Z^*({\boldsymbol q})}{E-E_n}$ where $Z({\boldsymbol q})$ is the Bethe-Salpeter (bound-state) wave function of the four-body bound state. We can thus obtain $Z({\boldsymbol q})$ by comparing the residue of both hand sides of the Skornyakov-Ter-Martirosian-type equation at $E=E_n$. Consequently, we obtain the following bound-state equation:
	\begin{widetext}
	\begin{align}
		Z({\boldsymbol q}')&=\int\frac{d^2l'}{(2\pi)^2}\int\frac{d^2l}{(2\pi)^2}
		\frac{\frac{\alpha}{\alpha+1}\chi\left({\boldsymbol l},\frac{\alpha}{2\alpha+1}{\boldsymbol q}'+{\boldsymbol l}'\right)
			\chi\left({\boldsymbol l},\frac{\alpha}{2\alpha+1}{\boldsymbol l}'+{\boldsymbol q}'\right)}
			{l^2+\frac{\alpha}{\alpha+1}\left[E+\frac{1}{\alpha}(q'^2+l'^2)+\frac{1}{\alpha+1}({\boldsymbol q}'+{\boldsymbol l}')^2\right]}\nonumber\\
			&\times \left[ -\frac{1}{32\pi^2}\frac{\alpha^2}{2\alpha+1}
			\left(E+\frac{3\alpha+1}{\alpha(2\alpha+1)}l'^2\right) 
			f\left(\frac{E+\frac{3\alpha+1}{\alpha(2\alpha+1)}l'^2}{\Lambda^2}\right)\right]^{-1}Z({\boldsymbol q}),\label{eq:bseq1}
	\end{align}
	\end{widetext}
	
	In solving Eq.~(\ref{eq:bseq1}), we first decompose $Z({\boldsymbol p})$ into different partial-wave sectors $Z({\boldsymbol q})=\sum_{n=-\infty}^{\infty} e^{in\theta}Z_n(q)$, where each $Z_n(q)$ is decoupled from another $Z_{n'}(q)$ due to the angular-momentum conservation. For each partial-wave sector, Eq.~(\ref{eq:bseq1}) can be solved analytically under the leading-logarithmic approximation \cite{nishida2013,moroz2014,nishida2017}, in which we assume that the dominant contribution in the integral on the right-hand side comes from the region of $\frac{E}{\Lambda^2}\ll \frac{q'^2}{\Lambda^2} \ll \frac{l'^2}{\Lambda^2} \ll 1$ and $\frac{E}{\Lambda^2}\ll \frac{l'^2}{\Lambda^2} \ll \frac{q'^2}{\Lambda^2} \ll 1$. We thus make an approximation in which the sum of $\frac{E}{\Lambda^2}+\frac{l'^2}{\Lambda^2}+\frac{q'^2}{\Lambda^2}$ in the integrand is replaced by $\frac{l'^2}{\Lambda^2}$ ($\frac{q'^2}{\Lambda^2}$) in the region of $q'<l'$ ($l'<q'$). Consequently, the bound-state equations of $Z_l(q)$ become
	\begin{align}
		&Z_0(\xi')=\frac{2(2\alpha+1)^2}{(\alpha+1)(3\alpha+1)}\nonumber\\
		&\hspace{10pt}\times\left[
			\int^{\xi'}_{\delta}d\xi Z_0(\xi)+\int^{\ln\frac{\Lambda^2}{E}}_{\xi'}d\xi\frac{\xi'}{\xi}Z_0(\xi)\label{eq:bseqZ0}
			\right],\\
		&Z_{|n|\geq1}(q)=0,\label{eq:bseqZl}
	\end{align}
	where the dimensionless variables $\xi=-\ln\left[\frac{3\alpha+1}{\alpha(2\alpha+1)}\frac{l'^2}{\Lambda^2}+\frac{E}{\Lambda^2}\right]$, $\xi'=-\ln\left[\frac{3\alpha+1}{\alpha(2\alpha+1)}\frac{q'^2}{\Lambda^2}+\frac{E}{\Lambda^2}\right]$ and $\delta$ is a non-universal parameter of ${\cal O}(1)$ that depends on a specific choice of the separable potential $\chi$. Equation~(\ref{eq:bseqZl}) shows that the higher partial-wave sectors cannot support bound states within the leading logarithmic approximation. \par
	Concerning the $s$-wave sector $l=0$, we can solve Eq.~(\ref{eq:bseqZ0}) by mapping the integral equation to a differential equation. By differentiating Eq.~(\ref{eq:bseqZ0}) twice with respect to $\xi'$, we obtain the following equation:
	\begin{align}
		\frac{d^2}{d\xi^2}Z_0(\xi)=-\frac{2(2\alpha+1)^2}{(\alpha+1)(3\alpha+1)}\frac{Z_0(\xi)}{\xi},\label{eq:dbseqZ0}
	\end{align}
	which is solved as
	\begin{align}
		&Z_0(\xi)=s\left[A\cdot  J_1(s)+B\cdot N_1(s)\right],\label{eq:soldbseq}\\
		&s:=\sqrt{\frac{8(2\alpha+1)^2}{(\alpha+1)(3\alpha+1)}\xi},
	\end{align}
	where $J_1$ is the Bessel function of the first kind, $N_1$ is the Neumann function and $A$ and $B$ are the constants of integration. 
	The solution Eq.~(\ref{eq:soldbseq}) is further restricted by the following two boundary conditions obtained by substiting $\xi'=\delta$ or $\xi'=\ln\frac{\Lambda^2}{E}$ into Eq.~(\ref{eq:bseqZ0}):
	\begin{align}
		& Z_0\left(\delta\right)=\frac{2(2\alpha+1)^2}{(\alpha+1)(3\alpha+1)}\int^{\ln\frac{\Lambda^2}{E}}_{\delta}d\xi\frac{\delta}{\xi}Z_0(\xi),\label{eq:bc1}\\
		& Z_0\left(\ln\frac{\Lambda^2}{E}\right)=\frac{2(2\alpha+1)^2}{(\alpha+1)(3\alpha+1)}\int^{\ln\frac{\Lambda^2}{E}}_{\delta}d\xi Z_0(\xi).\label{eq:bc2}
	\end{align}
	In the low-energy limit $\frac{E}{\Lambda^2}\rightarrow0$, the first boundary condition Eq.~(\ref{eq:bc1}) relates $A/B$ to the short-range non-universal parameter $\delta$ via
	\begin{align}
		&\frac{A}{B}=-\frac{\sigma\cdot N_1(\sigma)-\frac{\sigma^2}{2}\int_{\sigma}^{\infty}ds N_1(s)}
			{\sigma\cdot J_1(\sigma)-\frac{\sigma^2}{2}\int_{\sigma}^{\infty}ds J_1(s)},\label{eq:bcbswf1}\\
		&\sigma:=\sqrt{\frac{8(2\alpha+1)^2}{(\alpha+1)(3\alpha+1)}\delta}.
	\end{align}
	The other boundary condition Eq.~(\ref{eq:bc2}) determines the values of the binding energy $E$ via
	\begin{align}
		& \frac{A}{B}=-\frac{\theta\cdot N_1(\theta)-\frac{\theta^2}{2}N_2(\theta)}
				{\theta\cdot J_1(\theta)-\frac{\theta^2}{2}J_2(\theta)}
			\xrightarrow{\theta\rightarrow\infty} \tan\left(\frac{5\pi}{4}-\theta\right),\label{eq:bcbswf2}\\
		&\theta:=\sqrt{\frac{8(2\alpha+1)^2}{(\alpha+1)(3\alpha+1)}\ln\frac{\Lambda^2}{E}}.
	\end{align}
	It is straightforward to see that Eq.~(\ref{eq:bcbswf2}) in the low-energy limit $\theta\rightarrow\infty$ is satisfied only for $\theta=\theta^*+n\pi$ ($n\in{\mathbb Z}$) since the left-hand side is a $\theta$-independent constant. We thus obtain the quantized binding energy $E_n$ of the tetramers as
	\begin{align}
		E_n=\Lambda^2\exp\left[-\frac{(\alpha+1)(3\alpha+1)}{8(2\alpha+1)^2}(n\pi+\theta^*)^2\right],\label{eq:2cbEn}
	\end{align}
	or equivalently,
	\begin{align}	
		\sqrt{\ln\frac{\Lambda^2}{E_{n+1}}}-\sqrt{\ln\frac{\Lambda^2}{E_{n}}}=\pi\sqrt{\frac{(\alpha+1)(3\alpha+1)}{8(2\alpha+1)^2}},\label{eq:2cbEnpe}
	\end{align}
	which is nothing but the energy spectrum of the semisuper-Efimov effect.\par
	In Eq.~(\ref{eq:2cbEn}), the scaling parameter $\frac{(\alpha+1)(3\alpha+1)}{8(2\alpha+1)^2}$ is unexpectedly found to be stable under the variation of the mass-ratio $\alpha$. 
	In particular, major qualitative differences of the present result from Efimov and super-Efimov effects can be found in the extremely mass-imbalanced regimes $\alpha\rightarrow0$ and $\alpha\rightarrow\infty$: Compared to the Efimov and the super-Efimov effects, we observe the nonvanishing scaling parameter in the limit of $\alpha\rightarrow 0$ and the nondiverging scaling parameter in the limit of $\alpha\rightarrow\infty$.

	
	\subsubsection{Universality}\label{sec:2cbrg}
	\begin{figure}[t]
  		\begin{center}
  			 	\includegraphics[width=200pt]{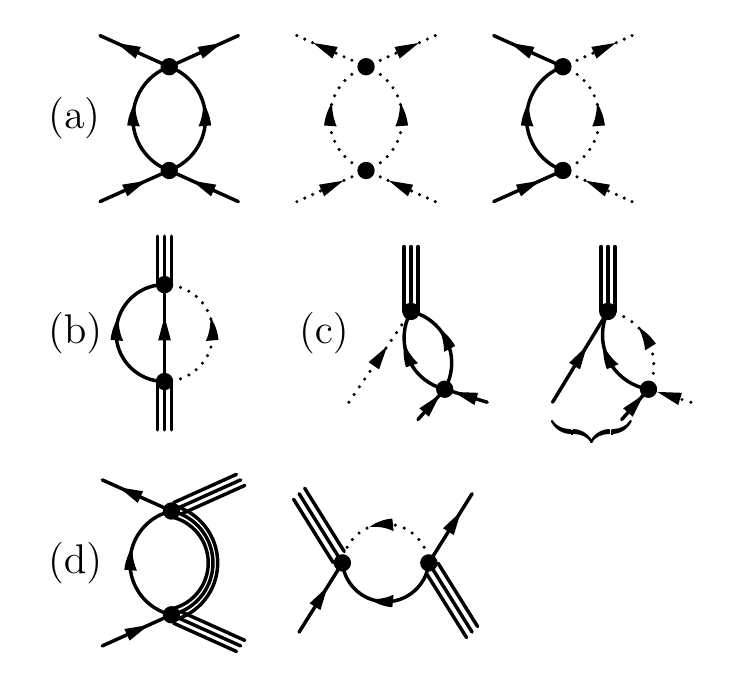}
				\caption{The Feynman diagrams that contribute to the renormalization-group equations of (a) the two-body coupling constants $g_A$, $g_B$ and $g_{AB}$. (b) the wave-function renormalization of the trimer field $\phi$. (c) the three-body coupling constant $h$. (d) the four-body coupling constant $v_4$. The curly bracket represents the symmetrization with respect to the indistinguishable bosons. }\label{fig:fig3}
  			\end{center}
	\end{figure}
	To discuss the universality of the analytically calculated binding energy Eq.~(\ref{eq:2cbEn}) in mass-imbalanced bosons, here we perform a renormalization-group analysis. We first deal with the situation with vanishing two-body interactions to reproduce Eq.~(\ref{eq:2cbEn}), and then we discuss the stability of the solution in presence of the two-body interactions. For this purpose, we consider the following effective-field theory by introducing an auxiliary field of a composite of three bosons:
	\begin{align}
		L&=\psi^{\dag}_A\left(i\partial_t+\frac{\nabla^2}{\alpha}\right)\psi_A
			+\psi^{\dag}_B\left(i\partial_t+\nabla^2\right)\psi_B \nonumber\\
			&+\phi^{\dag}\left(i\partial_t+\frac{\nabla^2}{2\alpha+1}-\epsilon_0\right)\phi
			+\frac{h}{2}\left(\phi^{\dag}\psi_A\psi_A\psi_B+\mbox{h.c.}\right) \nonumber\\
			&+v_4\phi^{\dag}\psi^{\dag}_A\psi_A\phi
			+v'_4\phi^{\dag}\psi^{\dag}_B\psi_B\phi
			+\frac{g_A}{4}\psi^{\dag}_A\psi^{\dag}_A\psi_A\psi_A\nonumber\\
			&+\frac{g_B}{4}\psi^{\dag}_B\psi^{\dag}_B\psi_B\psi_B
			+g_{AB}\psi^{\dag}_A\psi^{\dag}_B\psi_B\psi_A
			+\frac{g_{\phi}}{4}\phi^{\dag}\phi^{\dag}\phi\phi,\nonumber\\
	\end{align}
	where we introduce all the symmetry-preserving relevant couplings consisting of the two species $\psi_A$, $\psi_B$ of bosons and the trimer $\phi$. For the four-body problem of three $A$ bosons and a $B$ boson, the couplings $v'_4$ and $g_{\phi}$ are decoupled and do not enter the renormalization-group equations. To ensure that the system is at the three-body resonance, we tune $\epsilon_0$ at each renormalization-group energy scale of $\mu$ as
	\begin{align}
		\epsilon_0=-\frac{h^2\ln4}{32\pi^2}\frac{\alpha^2}{2\alpha+1}(\Lambda^2-\mu^2),
	\end{align}
	where $\Lambda$ is an intrinsic ultraviolet cutoff. Beta functions of renormalization-group equations are then given by coefficients of logarithmically divergent Feynman diagrams \cite{nishida2013, nishida2017} which are collected in Fig.~\ref{fig:fig3}. We thus obtain the following renormalization-group equations:
	\begin{align}
		&\frac{dg_A}{ds}=-\frac{\alpha g_A^2}{8\pi},\;\frac{dg_B}{ds}=-\frac{ g_B^2}{8\pi},\;\frac{dg_{AB}}{ds}=-\frac{g_{AB}^2}{2\pi}\frac{\alpha}{\alpha+1},\label{eq:RGEQ2cb2b}\\
		&\frac{dh}{ds}=-\frac{h^3}{32\pi^2}\frac{\alpha^2}{2\alpha+1}-\frac{\alpha g_A h}{8\pi}-\frac{g_{AB}h}{\pi}\frac{\alpha}{\alpha+1},\label{eq:RGEQ2cb3b}\\
		&\frac{dv_4}{ds}=-\frac{v_4 h^2}{16\pi^2}\frac{\alpha^2}{2\alpha+1}-\frac{v_4^2}{2\pi}\frac{\alpha(2\alpha+1)}{3\alpha+1}-\frac{h^2}{2\pi}\frac{\alpha}{\alpha+1},\label{eq:RGEQv_4}
	\end{align} 
	where $s:=\ln\Lambda/\mu$ is the renormalization time.\par
	To verify the result in Eq.~(\ref{eq:2cbEn}), we first consider the situation where two-body interactions are abscent: $g_A=g_B=g_{AB}=0$. In the situation, $h$ becomes
	\begin{align}
		h(s)^2=\left(\frac{1}{h(0)^2}+\frac{s}{16\pi^2}\frac{\alpha^2}{2\alpha+1}\right)^{-1},\label{eq:2bchrgsol}
	\end{align}
	which leads to $h(s)^2\rightarrow \left(\frac{s}{16\pi^2}\frac{\alpha^2}{2\alpha+1}\right)^{-1}$ in the low-energy limit $s\rightarrow\infty$. By substituting this expression into Eq.~(\ref{eq:RGEQv_4}), we obtain
	\begin{align}
		v_4=\frac{4\pi}{\alpha}\sqrt{\frac{3\alpha+1}{\alpha+1}}\frac{1}{\sqrt{s}}
			\tan\left(-\sqrt{\frac{8(2\alpha+1)^2}{(\alpha+1)(3\alpha+1)}}\sqrt{2s}+C\right),\nonumber\\
		{}\label{eq:2cbsolv4}
	\end{align}
	where $C$ is a non-universal constant that depends on the initial value of $v_4(0)$ (see Appendex~\ref{sec:app1} for a detailed calculation).
	
	Since a divergence of a coupling constant is a fingerprint of an emergence of a bound state, we assign $\mu=\mu_n$ ($n\in {\mathbb Z}$) at which $v_4\left(\ln \frac{\Lambda}{\mu}\right)$ diverges. Due to the periodic nature of $\sqrt{s}v_4(s)$ with respect to $\sqrt{2s}=\sqrt{\ln\frac{\Lambda^2}{\mu^2}}$, we obtain
	\begin{align}
		\sqrt{\ln\frac{\Lambda^2}{\mu_{n+1}^2}}-\sqrt{\ln\frac{\Lambda^2}{\mu_{n}^2}}=\pi\sqrt{\frac{(\alpha+1)(3\alpha+1)}{8(2\alpha+1)^2}},
	\end{align}
	 which is in perfect agreement with the period of the obtained energy spectrum Eq.~(\ref{eq:2cbEnpe}) of the semisuper-Efimov effect.  
	 
	To discuss the stability of the semisuper-Efimov states against the two-body interaction, we then consider the situation where the two-body couplings $g_A$, $g_B$ and $g_{AB}$ take on nonzero values. Although Nishida discuss in Ref.~\cite{nishida2017} that the semisuper-Efimov effect vanishes if we introduce a two-body interaction, we find it is not the case. Even in presence of two-body interactions, we obtain the same solution as Eq.~(\ref{eq:2cbsolv4}) under the following conditions:
	\begin{align}
		&\left|\frac{1}{g_A(0)}\right|\gg\frac{\alpha s}{8\pi},\;\left|\frac{1}{g_{AB}(0)}\right|\gg\frac{s}{2\pi}\frac{\alpha}{\alpha+1},
		\label{eq:2cbcond1}\\
		&h(0)^2\gg 32\pi^2\frac{2\alpha+1}{\alpha^2}
			\left(\frac{\alpha}{\alpha+1}\frac{g_{AB}(0)}{\pi}+\frac{\alpha g_A(0)}{8\pi}\right),\label{eq:2cbcond2}\\
		&s\gg \frac{16\pi^2(2\alpha+1)}{\alpha^2 h(0)^2}.\label{eq:2cbcond3}
	\end{align}
	In Appendix~\ref{sec:app2}, we explicitly derive Eq.~\eqref{eq:2cbsolv4} under the conditions Eqs.~\eqref{eq:2cbcond1}, \eqref{eq:2cbcond2} and \eqref{eq:2cbcond3}; here we discuss physical significance of these conditions.
	Firstly, in Eq.~(\ref{eq:2cbcond1}), the inverse two-body coupling constants $1/|g_A(0)|$ and $1/|g_{AB}(0)|$ determine the upper bound of $s=\ln\Lambda/\mu$ (the lower bound of $\mu$) below which the semisuper-Efimov states are present, i.e., $1/|g_A(0)|$ and $1/|g_{AB}(0)|$ serve as infrared cutoffs of the spectrum of the semisuper-Efimov tetramers. Physically, the result suggests that an arbitrarily large quantum halo is prohibited by the two-body interactions and the possible size of the largest semisuper-Efimov tetramer is given by $\min\{\frac{1}{\Lambda}e^{\frac{8\pi}{\alpha|g_A(0)|}},\frac{1}{\Lambda}e^{\frac{2\pi(\alpha+1)}{\alpha|g_{AB}(0)|}}\}$. In other words, the minimum value of the binding energy of the semisuper-Efimov tetramer is given by $\max\{\Lambda^2 e^{-\frac{16\pi}{\alpha|g_A(0)|}},\Lambda^2 e^{-\frac{4\pi(\alpha+1)}{\alpha|g_{AB}(0)|}}\}$. It is reasonable to consider that semisuper-Efimov tetramers are absent below the energy scales $\Lambda^2 e^{-\frac{16\pi}{\alpha|g_A(0)|}}$ and $\Lambda^2 e^{-\frac{4\pi(\alpha+1)}{\alpha|g_{AB}(0)|}}$ which are the binding energies of two-body bound states. The second condition Eq.~(\ref{eq:2cbcond2}) means that the three-body coupling constant $h(0)$ must be sufficiently larger than the two-body couplings $g_A(0)$ and $g_{AB}(0)$, so that the resonant three-body interaction overwhelms the weak two-body interactions (quantum fluctuation is dominated by the loop corrections originating in the three-body interaction). The final condition Eq.~(\ref{eq:2cbcond3}) ensures that the semisuper-Efimov effect occur at a sufficiently low-energy regime where the only non-vanishing energy scale is the intrinsic ultraviolet cutoff $\Lambda$. We note that the three conditions Eqs.~(\ref{eq:2cbcond1}), (\ref{eq:2cbcond2}) and (\ref{eq:2cbcond3}) are compatible with each other.
	
	In conclusion, we verify the analytically obtained energy spectrum Eq.~(\ref{eq:2cbEn}) by comparing the spectrum with a solution Eq.~(\ref{eq:2cbsolv4}) of a renormalization-group equation. Moreover, by clarifying the conditions under which the solution Eq.~(\ref{eq:2cbsolv4}) is stable, we show the stability of the semisuper-Efimov effect in presence of weak two-body interactions.

	\subsection{Three-component bosons}\label{sec:threedistinguisable}
	As another system to investigate, we here consider a three-component bosons in which two bosons of the three species have an identical mass. Since the calculation procedure is almost same as that performed for two-component bosons, we here present the results.
	The Hamiltonian we consider is the following:
	\begin{widetext}
	\begin{align}
		&H=\int\frac{d^2q}{(2\pi)^2}\frac{{\boldsymbol q}^2}{\alpha}\psi^{\dag}_{\uparrow,{\boldsymbol q}}\psi_{\uparrow,{\boldsymbol q}}
		+\int\frac{d^2q}{(2\pi)^2}\frac{{\boldsymbol q}^2}{\alpha}\psi^{\dag}_{\downarrow,{\boldsymbol q}}\psi_{\downarrow,{\boldsymbol q}}
		+\int\frac{d^2q}{(2\pi)^2}{\boldsymbol q}^2\psi^{\dag}_{B,{\boldsymbol q}}\psi_{B,{\boldsymbol q}}
		-\lambda\int\frac{d^2Q d^2q'_x d^2q'_y d^2q_x d^2q_y}{(2\pi)^{10}}
		\chi({\boldsymbol q}'_x,{\boldsymbol q}'_y)\chi({\boldsymbol q}_x,{\boldsymbol q}_y)\nonumber\\
		&\hspace{100pt}\times
		\psi^{\dag}_{\uparrow, \frac{ \alpha {\boldsymbol Q}}{2\alpha+1}+{\boldsymbol q}'_y}
		\psi^{\dag}_{\downarrow, \frac{ \alpha {\boldsymbol Q}}{2\alpha+1}-\frac{\alpha{\boldsymbol q}'_y}{\alpha+1}+{\boldsymbol q}'_x}
		\psi^{\dag}_{B,\frac{\boldsymbol Q}{2\alpha+1}-\frac{{\boldsymbol q}'_y}{\alpha+1}-{\boldsymbol q}'_x}
		\psi_{B,\frac{\boldsymbol Q}{2\alpha+1}-\frac{{\boldsymbol q}_y}{\alpha+1}-{\boldsymbol q}_x}
		\psi_{\downarrow, \frac{ \alpha {\boldsymbol Q}}{2\alpha+1}-\frac{\alpha{\boldsymbol q}_y}{\alpha+1}+{\boldsymbol q}_x}
		\psi_{\uparrow, \frac{ \alpha {\boldsymbol Q}}{2\alpha+1}+{\boldsymbol q}_y}.
		\label{eq:H_3bosons}
	\end{align}
	\end{widetext}
	Here we employ the units $\hbar=2m=1$, where $m$ is the mass of a boson $\psi_B$. The annihilation (creation) operators $\psi_{\uparrow}$ and $\psi_{\downarrow}$ ($\psi_{\uparrow}^{\dag}$ and $\psi_{\downarrow}^{\dag}$) represent two species of bosons who have an identical mass of $\alpha/2$. To see the effects of the quantum statistics of the particles, here we assume that the three-body interaction among the three species of bosons is tuned to its resonance, and the other three-body interactions are not. In the three-body interaction, we employ the separable interaction potential $\chi$ in Eq.~(\ref{eq:regulator}). \par
	
	\begin{figure}[t]
  		\begin{center}
  			 \includegraphics[width=240pt]{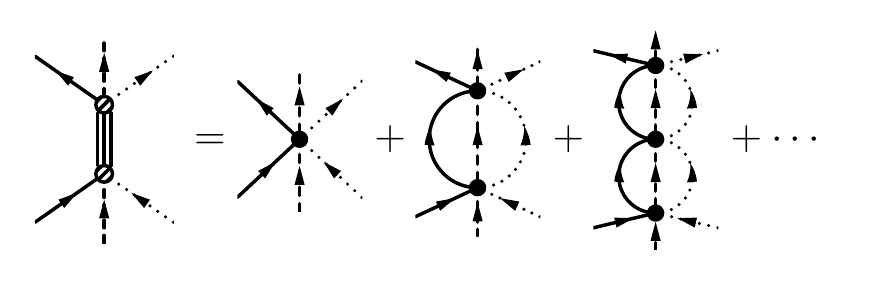}
			\caption{Diagrammatic expression of the three-body $T$-matrix $T^{(3)}$ for three-component bosons, where $T^{(3)}$ is represented by the triple lines on the left-hand side. The shaded circle on the left-hand side represents the ultraviolet-regulator function $\chi$ which is not renormalized by quantum fluctuations. Here the solid, the dashed and the dotted lines represent the propagator of the three species of bosons $\psi_{\uparrow}$, $\psi_{\downarrow}$ and $\psi_B$, respectively.}\label{fig:fig4}
  		\end{center}
	\end{figure}
	\begin{figure}[t]
  		\begin{center}
  			 \includegraphics[width=200pt]{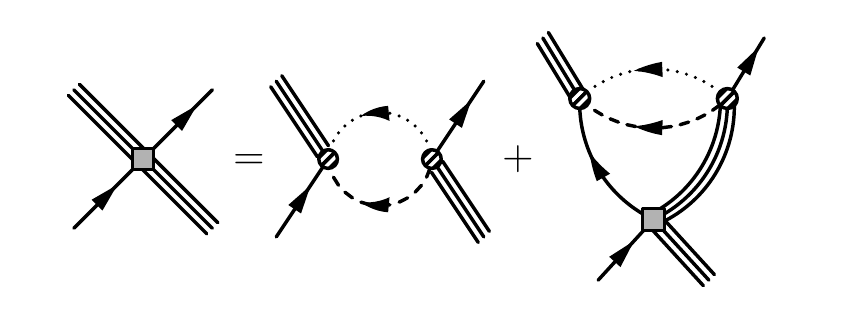}
			\caption{The Skornyakov-Ter-Martirosian-type integral equation of the four-body $T$-matrix $T^{(4)}$ for three-component bosons, where $T^{(4)}$ is represented by the shaded square on the left-hand side. The equation can be obtained by summing up the ladder-type Feynman diagrams that refer to the two-particle exchange processes.}\label{fig:fig5}
  		\end{center}
	\end{figure}
	
	For the system, we perform the same analysis as two-component mass-imbalanced bosons, i.e. we solve the Bethe-Salpeter equation analytically. Since the procedure employed here is same as that of Sec.~\ref{sec:2cbma}, we here just list up the results. Firstly, the three-body $T$-matrix $T^{(3)}$ of the three-distinguishable bosons is obtained by summing up the ladder-type Feynman diagrams depicted in Fig.~\ref{fig:fig4}. Consequently, we obtain
	\begin{align}
		T^{(3)}(p^0,{\boldsymbol p}; {\boldsymbol q}'_x,{\boldsymbol q}'_y, {\boldsymbol q}_x,{\boldsymbol q}_y)
		=\frac{\chi({\boldsymbol q}'_x,{\boldsymbol q}'_y)\chi({\boldsymbol q}_x,{\boldsymbol q}_y)}
			{-\frac{1}{g}-\frac{1}{16\pi^2}\frac{\alpha^2\epsilon}{2\alpha+1}f\left(\frac{\epsilon}{\Lambda^2}\right)},\label{eq:3cbT_3}
	\end{align}
	where the function $f$ is given in Eq.~(\ref{eq:funcf}). So that the system is on its three-body resonance, we renormalize the three-body coupling constant $\lambda$ as $\frac{1}{\lambda}-\frac{1}{16\pi^2}\frac{\alpha^2}{2\alpha+1}\Lambda^2=0$. \par
	Using the three-body $T$-matrix, we then turn to the four-body sector where we consider the scattering of the three-distinguishable bosons and an additional boson of $\psi_{\uparrow}$. The four-body Bethe-Salpeter equation is then obtained by seeing a pole structure of the Skornyakov-Ter-Martirosian-type integral equation depicted in Fig.~\ref{fig:fig5}. Corresponding to Eq.~\eqref{eq:bseqZ0}, we finally obtain the nonvanishing Bethe-Salpeter equation for the $s$-wave sector $Z_0$ as
	\begin{align}
		&Z_0(\xi')=\frac{(2\alpha+1)^2}{(\alpha+1)(3\alpha+1)}\nonumber\\
		&\hspace{10pt}\times\left[
			\int^{\xi'}_{\delta}d\xi Z_0(\xi)+\int^{\ln\frac{\Lambda^2}{E}}_{\xi'}d\xi\frac{\xi'}{\xi}Z_0(\xi)
			\right].\label{eq:3cbbseqZ0}
	\end{align}
	By following the same procedure as Sec.~\ref{sec:2cbma}, we finally arrive at the following energy spectrum of tetramers:
	\begin{align}
		&E_n=\Lambda^2\exp\left[-\frac{(\alpha+1)(3\alpha+1)}{4(2\alpha+1)^2}(n\pi+\theta^*)^2\right],\label{eq:3cbEn}\\
		&\sqrt{\ln\frac{\Lambda^2}{E_{n+1}}}-\sqrt{\ln\frac{\Lambda^2}{E_{n}}}=\pi\sqrt{\frac{(\alpha+1)(3\alpha+1)}{4(2\alpha+1)^2}},\label{eq:3cbEnpe}
	\end{align}
	where $\theta^*$ is a nonuniversal constant. \par
	Again we find an emergence of the semisuper-Efimov effect with a mass-ratio-dependent scaling parameter. In particular, presence of the tetramer states in extremely mass-imbalanced situations ($\alpha\gg1$ and $\alpha\ll1$) is observed.

	\section{Particles with linear-dispersion relation in 1D}\label{sec:1dlinear}
	To further extend the universality of the semisuper-Efimov effect, we here consider a one-dimensional system of identical bosons that have a linear-dispersion relation. 
	\subsection{model analysis}
\begin{figure}[t]
  		\begin{center}
  			 \includegraphics[width=240pt]{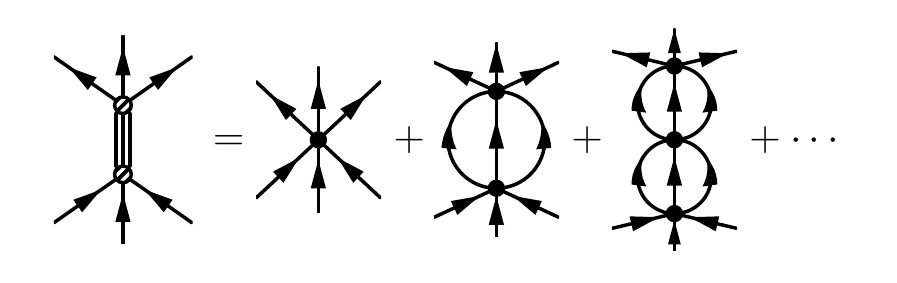}
			\caption{Diagrammatic expression of the three-body $T$-matrix, which is represented by the triple lines on the left-hand side. The shaded circle on the left-hand side represents the ultraviolet-regulator function $\chi$ which is not renormalized by quantum fluctuations. Here the solid line represents the propagator of the bosonic particle which has the linear dispersion relation.}\label{fig:fig6}
  		\end{center}
	\end{figure}

	We consider the following model Hamiltonian:
	\begin{widetext}
	\begin{align}
		H=\int\frac{dq}{2\pi}|q|\psi^{\dag}_{q}\psi_{q}
		-\frac{\lambda}{(3!)^2}\int\frac{dQ dq'_x dq'_y dq_x dq_y}{(2\pi)^{5}}
		\chi(q'_x,q'_y)\chi(q_x,q_y)\nonumber\\
		\times
		\psi^{\dag}_{\frac{Q}{2}+q'_y}
		\psi^{\dag}_{\frac{Q}{4}-\frac{q'_y}{2}+q'_x}
		\psi^{\dag}_{\frac{Q}{4}-\frac{q'_y}{2}-q'_x}
		\psi_{\frac{Q}{4}-\frac{q_y}{2}-q_x}
		\psi_{\frac{Q}{4}-\frac{q_y}{2}+q_x}
		\psi_{\frac{Q}{2}+q_y},
		\label{eq:H_1Dbosons}
	\end{align}
	where $\psi^{\dag}$ ($\psi$) is the creation (annihilation) operator of a boson. Here we employ the units $\hbar=v=1$, where $v$ is the velocity of the linearly dispersing boson. As a separable potential $\chi$, we choose the following sharp-cutoff function:
	\begin{align}
		\chi(q_x,q_y)=\Theta(\Lambda-q_x)\Theta(2\Lambda-q_y),
	\end{align}
	where $\Theta$ is the Heavyside unit-step function. 
	
	In the system, we first consider the three-body sector in which the three-body $T$-matrix $T^{(3)}$ is the summation of the ladder-type Feynman diagrams depicted in Fig.~\ref{fig:fig6}. Consequently, we obtain
	\begin{align}
		& T^{(3)}(p^0,p;q_x',q_y',q_x,q_y)=-\chi(q_x',q_y')\Gamma(p^0,p)\chi(q_x,q_y),\\
		&-\Gamma(p^0,p)=\frac{-6}
		{\frac{6}{\lambda}+\frac{4\ln2+1}{2\pi^2}\Lambda+\frac{3ip^0}{8\pi^2}\ln\frac{ip^0+|p|}{4\Lambda}
		+\frac{ip^0}{4\pi^2}\ln\frac{4}{e}-\frac{3}{2}|p|+\frac{p^2}{2}\frac{1}{ip^0+|p|}}.\label{eq:1db3bt}
	\end{align}
	\end{widetext}
	
	 \begin{figure}[t]
  		\begin{center}
  			 	\includegraphics[width=200pt]{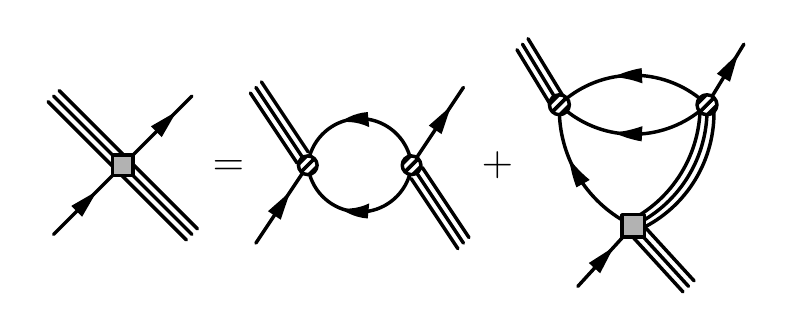}
				\caption{Skornyakov-Ter-Martirosian-type linear integral equation of the four-body $T$-matrix $T^{(4)}$ which is represented by the shaded square on the left-hand side. The equation can be obtained by summing up the ladder-type Feynman diagrams consisting of the two-particle exchange processes. We note that the second term on the right-hand side produces an additional symmetry factor 1/2 due to the exchange of two internal lines.}\label{fig:fig7}
  			\end{center}
	\end{figure}
	So that the three-body $T$-matrix $T^{(3)}$ is on its resonance, we renormalize the three-body coupling constant $\lambda$ as $\frac{6}{\lambda}+\frac{4\ln2+1}{2\pi^2}\Lambda=0$. We note that the dominant term in Eq.~(\ref{eq:1db3bt}) is the logarithmic function $\frac{3ip^0}{8\pi^2}\ln\frac{ip^0+|p|}{4\Lambda}$ in a sufficiently low-energy regime $p^0,|p|\ll\Lambda$. \par
	As discussed in Sec.~\ref{sec:2cbma}, the bound-state equation of the four-body sector is obtained by comparing the residue of the Skornyiakov-Ter-Martirosian-type integral equation depicted in Fig.~\ref{fig:fig7}. We note that there are two partial wave sectors in 1D labeled by the the parity quantum number. The bound-state equation with the binding energy $E$ can be obtained as
	\begin{align}
		Z(q)=\int_{0}^{\infty} dl \ln\left[\frac{2\Lambda}{E+2q+2l}\frac{2\Lambda}{E+q+l+|q-l|}\right]\nonumber\\
		\times\frac{-2Z(l)}{(E+l)\ln\frac{E+2l}{\Lambda}},
	\end{align}
	for the even-parity ($Z(-q)=Z(q)$) sector and 
	\begin{align}
		Z(q)=\int_{0}^{\infty} dl \ln\left[\frac{E+q+l+|q-l|}{E+q+l+|q+l|}\right]\frac{-2Z(l)}{(E+l)\ln\frac{E+2l}{\Lambda}},\nonumber\\
	\end{align}
	for the odd-parity ($Z(-q)=-Z(q)$) sector. For the odd parity sector, we find no bound state. Concerning the even-parity sector, we employ the leading-logarithmic approximation together with the change of variables $\xi:=\ln\frac{\Lambda}{E+2l}$ and $\xi':=\ln\frac{\Lambda}{E+2q}$. Then we have
	\begin{align}
		\psi(\xi')=4\int_{\delta}^{\xi'}d\xi Z(\xi)+4\int_{\xi'}^{\ln\frac{\Lambda}{E}}d\xi \frac{\xi'}{\xi}Z(\xi),
	\end{align}
	Similarly to the solution of Eq.~\eqref{eq:bseqZ0}, we obtain the following energy spectrum of tetramers:
	\begin{align}
		&E_n=\Lambda \exp\left[-\frac{1}{16}(n\pi+\theta^*)^2\right],\label{eq:1dbEn}\\
		&\sqrt{\ln\frac{\Lambda}{E_{n+1}}}-\sqrt{\ln\frac{\Lambda}{E_{n}}}=\frac{\pi}{4},\label{eq:1dbEnpe}
	\end{align}
	where $\theta^*$ is a nonuniversal constant determined by short-range details of a three-body interaction. We note that in Eqs.~\eqref{eq:1dbEn} and \eqref{eq:1dbEnpe}, a momenta and an energy have the same dimension due to the linearity of the dispersion relation. We thus show that the semisuper-Efimov effect occurs even in one dimension if a particle exhibits the linear dispersion relation.


	\subsection{universality}
	We here discuss the universality of the results by following the same procedure as Sec.~\ref{sec:2cbrg}. To perform a renormalization-group analysis, we consider an effective field theory with the following Action:
	\begin{widetext}
	\begin{align}
		&S=\int\frac{d^2q}{(2\pi)^2}\psi^{\dag}(q)(iq^0+|q|)\psi(q)+\int\frac{d^2q}{(2\pi)^2}\phi^{\dag}(q)(iq^0-\epsilon)\phi(q)\nonumber\\
		&+\int d^2x\frac{g_2}{4}\psi^{\dag}(x)\psi^{\dag}(x)\psi(x)\psi(x)
		+\frac{h}{6}\left[\phi^{\dag}(x)\psi(x)\psi(x)\psi(x)+\mbox{h.c.}\right]
		+g_4\phi^{\dag}(x)\psi^{\dag}(x)\psi(x)\phi(x),
	\end{align}
	\end{widetext}
	where $\psi$ and $\phi$ ($\psi^{\dag}$ and $\phi^{\dag}$) are the annihilation (creation) operators of a bosons and a trimer, respectively. Here we list up all the symmetry-preserving relevant terms up to the four-body sector, since the higher-body sectors are decoupled does not enter the renormalization-group equations of the lower-body sectors. So that the system is on its three-body resonance, we here tune the parameter $\epsilon=-\frac{h^2}{12\pi^2}(4\ln2+1)(\Lambda-\mu)$ at each renormalization-group scale $\mu$.
	
\begin{figure}[t]
  		\begin{center}
  			 \includegraphics[width=200pt]{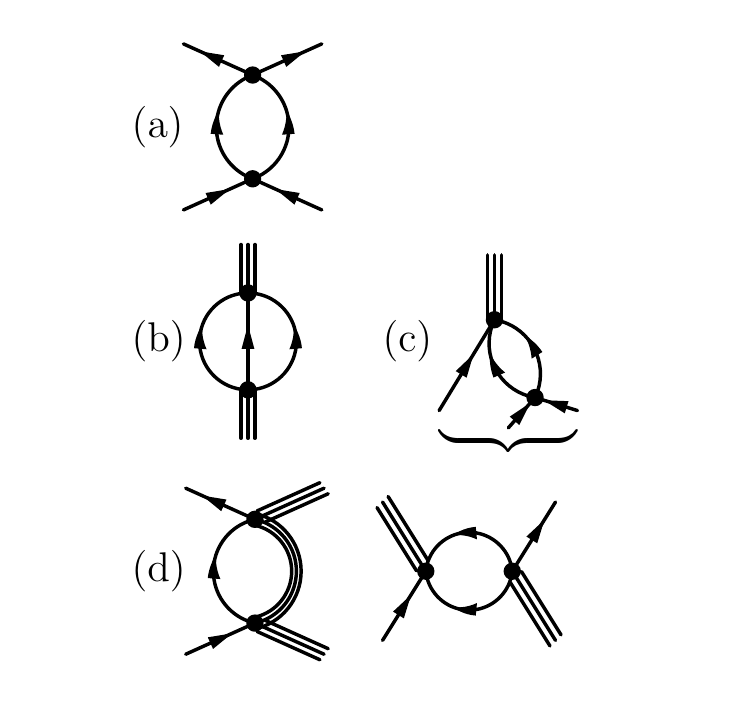}
			\caption{The Feynman diagrams that contribute to the renormalization-group equations of (a) the two-body coupling constant $g_2$. (b) the wave-function renormalization of the trimer field $\phi$. (c) the three-body coupling constant $h$. (d) the four-body coupling constant $v_4$. }\label{fig:fig8}
  		\end{center}
	\end{figure}

	A beta function of a renormalization-group equation is obtained by summing up coefficients of logarithmically diverging Feynman diagrams which we collect in Fig.~\ref{fig:fig8}. Consequently we obtain the following renormalization-group equations:
	\begin{align}
		&\frac{dg_2}{ds}=-\frac{g_2^2}{4\pi},\\
		&\frac{dh}{ds}=\frac{h^3}{32\pi^2}-\frac{3g_2h}{4\pi},\\
		&\frac{dg_4}{ds}=\frac{h^2g_4}{16\pi^2}-\frac{g_4^2}{\pi}-\frac{h^2}{4\pi},\label{eq:1db4bc}
	\end{align}
	where we have introduced the renormalization-group time $s=\ln\Lambda/\mu$ similarly to Sec.~\ref{sec:2cbrg}.
	
	In absence of the two-body interaction $g_2(0)=0$, the three-body coupling constant $h$ becomes
	\begin{align}
		h(s)=\frac{1}{h(0)^{-2}-\frac{s}{16\pi^2}}\xrightarrow[]{s\rightarrow \infty} \frac{-16\pi^2}{s}.
	\end{align}
	By substituting this expression into Eq.~\eqref{eq:1db4bc}, we obtain
	\begin{align}
		g_4(s)=\frac{1}{\sqrt{s}}\cot\left(4\sqrt{s}+C\right),\label{eq:1db4bcs}
	\end{align}
	where $C$ is a constant of integration. As discussed in Sec.~\ref{sec:2cbrg}, a divergence of the four-body coupling constant $g_4$ is a fingerprint of an emergence of a tetramer. Therefore, we assign $\mu=\mu_n$ ($n\in\mathbb{Z}$) where $g_4\left(\frac{\Lambda}{\mu}\right)$ diverges. Due to the periodic nature Eq.~\eqref{eq:1db4bcs} of $\sqrt{s}g_4(s)$ with respect to $4\sqrt{s}$, we find
	\begin{align}
		\sqrt{\ln\frac{\Lambda}{\mu_{n+1}}}-\sqrt{\ln\frac{\Lambda}{\mu_{n}}}=\frac{\pi}{4},\label{eq:1db4bcpd}
	\end{align}
	 which is in perfect agreement with the energy spectrum Eq.~\eqref{eq:1dbEnpe}.
	 
	 Similarly to Sec.~\ref{sec:2cbrg} (see also Appendix~\ref{sec:app2}), the solution Eq.~\eqref{eq:1db4bcs} is stable even if we introduce a finite two-body interaction $g_2(0)$. Specifically, the same solution Eq.~\eqref{eq:1db4bcs} for the four-body coupling constant $g_4$ under the following conditions:
	 \begin{align}
	 	&\left|\frac{1}{g_2(0)}\right|\gg \frac{s}{4\pi},\label{eq:1dbcond1}\\
		&h(0)^2\gg 24\pi g_2(0),\label{eq:1dbcond2}\\
		&s\gg \frac{32\pi^2}{h(0)^2}.\label{eq:1dbcond3}
	 \end{align}
	 
	 In conclusion, we obtain a consistent result Eq.~\eqref{eq:1db4bcpd} with the energy spectrum Eq.~\eqref{eq:1dbEnpe} of the semisuper-Efimov effect. Furthermore, we argue that a weak two-body interaction sustains the semisuper-Efimov tetramers in the parameter region given by Eqs.~\eqref{eq:1dbcond1}, \eqref{eq:1dbcond2} and \eqref{eq:1dbcond3}.
	 
	\section{summary and discussion}\label{sec:summary}
	We here summarize the discussion in the main text.  Firstly, for mass-imbalanced bosons in 2D, we demonstrate an emergence of the semisuper-Efimov effect and derive mass-ratio dependent scaling factor. The scaling factor is unexpectedly found to be stable under the variation of the mass-ratio and, in particular, we find that the scaling factor remains finite in extremely mass-imbalanced situation where the mass ratio takes 0 or $\infty$. This is in clear contrast with the Efimov and the super-Efimov effects where the scaling factors vanish or diverge for an extreme-mass imbalance. A renormalization-group analysis is then performed and a consistent renormalization-group flow with the semisuper-Efimov effect is obtained. Furthermore, by introducing a finite two-body interaction, we show that the semisuper-Efimov effect is stable even in the presence of a weak two-body interaction. The semisuper-Efimov effect is then shown to emerge also in the system of three-component bosons. Finally, we liberate the semisuper-Efimov effect from 2D, by considering bosonic particles with linear-dispersion relation in 1D.
	
	Because of the presence of the semisuper-Efimov effect in an extremely mass-imbalanced situation, the effect might be of relevance in impurity problems where an impurity has a large inertial mass. For example, consider a system of identical bosons with a spatially localized external potential such as a narrow square-well potential. If an interaction between two identical bosons is tuned to be on its resonance {\it only inside} the potential well, the two-body interaction between identical bosons can effectively be regarded as a three-body interaction between two identical bosons and the external potential: 
	\begin{align}
		V({\bf r}_A,{\bf r}_B)\propto\delta({\bf r}_P-{\bf r}_A)\delta({\bf r}_P-{\bf r}_B),
	\end{align}
	where ${\bf r}_A$ and ${\bf r}_B$ refers to the positions of identical bosons and ${\bf r}_P$ is the position of the external potential. Experimentally, a spatial control of interaction is already realized in Ref.~\cite{clark2015}. If the optical control of interaction introduced in Ref.~\cite{clark2015} is implemented in a system with single-atom resolution, the above effective short-range three-body interaction might be realized. Analogous to the Efimov effect, a resonance of the atomic loss will be a fingerprint of the semisuper-Efimov effect.

	\begin{acknowledgements}
		The author acknowledges Takeshi Fukuhara for helpful discussions. The author is supported by RIKEN Special Postdoctoral Researcher Program.
	\end{acknowledgements}

	\appendix
	\section{Solution of the four-body renormalzation-group equation}\label{sec:app1}
	We here sketch the derivation of Eq.~(\ref{eq:2cbsolv4}), which is the solution of the four-body renormalization-group equation Eq.~(\ref{eq:RGEQv_4}) in the situation of vanishing two-body coupling constants $g_A=g_B=g_{AB}=0$. Since the same procedure applies to systems of three-component bosons and bosons in 1D, we here demonstrate the calculation in the two-component mass-imbalanced bosons. By substituting the asymptotic form $h(s)\xrightarrow{s\rightarrow\infty} \frac{16\pi^2}{s}\frac{2\alpha+1}{\alpha^2}$ of the three-body coupling constant into Eq.~(\ref{eq:RGEQv_4}), we obtain
	\begin{align}
		\frac{dv_4}{ds}=-\frac{v_4}{s}
			-\frac{v_4^2}{2\pi}\frac{\alpha(2\alpha+1)}{3\alpha+1}
			-\frac{8\pi}{s}\frac{2\alpha+1}{\alpha(\alpha+1)}.
	\end{align}
	Since we expect that the solution $v_4(s)$ is periodic with respect to the variable $\sqrt{s}$, we introduce convenient variables $s=:t^2$ and $g_4:=t v_4$. Consequently, we obtain
	\begin{align}
		\frac{dg_4}{dt}=-\frac{g_4}{t}
		-\frac{g_4^2}{\pi}\frac{\alpha(2\alpha+1)}{3\alpha+1}
		-16\pi\frac{2\alpha+1}{\alpha(\alpha+1)}.\label{eq:2cbrgg4app}
	\end{align}
	In solving the equation, we assume that the first term on the right-hand side does not provide a dominant contribution in the low-energy limit $t\gg1$ and the term will be neglected hereafter. The assumption is justified by substituting the obtained solution of Eq.~(\ref{eq:2cbsolv4}) into Eq.~(\ref{eq:2cbrgg4app}). Namely, at sufficiently low-energy $t\gg1$, the first term on the right-hand side of Eq.~(\ref{eq:2cbrgg4app}) is $1/t$-times smaller than the rest two terms. We thus arrive at the following simple equation:
	\begin{align}
		-\frac{dg_4}{\frac{g_4^2}{\pi}\frac{\alpha(2\alpha+1)}{3\alpha+1}
		+16\pi\frac{2\alpha+1}{\alpha(\alpha+1)}}=dt,
	\end{align}
	which can be easily integrated to
	\begin{align}
		\arctan\frac{g_4(t)}{\frac{4\pi}{\alpha}\sqrt{\frac{3\alpha+1}{\alpha+1}}}
		=-4\sqrt{\frac{(2\alpha+1)^2}{(\alpha+1)(3\alpha+1)}}t+{\rm const.}\nonumber\\
		{}
	\end{align}
	We thus obtain Eq.~(\ref{eq:2cbsolv4}). \par
	As a complement, we note that an exact solution of Eq.~(\ref{eq:2cbrgg4app}) is given by
	\begin{widetext}
	\begin{align}
		g_4(t)=-\frac{4\pi}{\alpha}\sqrt{\frac{3\alpha+1}{\alpha+1}}
		\frac{J_1\left(4\sqrt{\frac{(2\alpha+1)^2}{(\alpha+1)(3\alpha+1)}}t\right)
			+N_1\left(4\sqrt{\frac{(2\alpha+1)^2}{(\alpha+1)(3\alpha+1)}}t\right)A}
		{J_0\left(4\sqrt{\frac{(2\alpha+1)^2}{(\alpha+1)(3\alpha+1)}}t\right)
			+N_0\left(4\sqrt{\frac{(2\alpha+1)^2}{(\alpha+1)(3\alpha+1)}}t\right)A},
	\end{align}
	\end{widetext}
	where $A$ is a constant. From the solution, we can also obtain Eq.~(\ref{eq:2cbsolv4}) by using asymptotic forms of the Bessel's functions $J_{\nu}$ and $N_{\nu}$.
	 
	\section{Effects of two-body interaction}\label{sec:app2}
	In Sec.~\ref{sec:2cbrg}, we discuss an emergence of the semisuper Efimov states in presence of two-body interactions. Here we show the derivation of Eq.~(\ref{eq:2cbsolv4}) under the conditions Eqs.~(\ref{eq:2cbcond1}), (\ref{eq:2cbcond2}) and (\ref{eq:2cbcond3}). Since the same discussion applies to systems of three-component bosons and bosons in 1D, we here consider the system of two-component mass-imbalanced bosons.
	Firstly, under the condition of Eq.~(\ref{eq:2cbcond1}), the solutions of the renormalization-group equations Eq.~(\ref{eq:RGEQ2cb2b}) of the two-body sector become
	\begin{align}
		&g_A(s)=\left(\frac{1}{g_A(0)}+\frac{\alpha s}{8\pi}\right)^{-1}\simeq g_A(0),\\
		&g_{AB}(s)=\left(\frac{1}{g_{AB}(0)}+\frac{\alpha s}{8\pi}\right)^{-1}\simeq g_{AB}(0).
	\end{align}
	By substituting these solutions into the renormalization-group equation Eq.~(\ref{eq:RGEQ2cb3b}) of the three-body coupling constant $h(s)$, we obtain
	\begin{align}
		&h(s)^2=\frac{\frac{16\pi^2C(2\alpha+1)h(0)^2}{\alpha^2 h(0)^2+16\pi^2C(2\alpha+1)}}
			{e^{Cs}-\frac{\alpha^2h(0)^2}{\alpha^2 h(0)^2+16\pi^2C(2\alpha+1)}},\\
		&C:=\frac{\alpha g_{A}(0)}{4\pi}+\frac{2\alpha}{\alpha+1}\frac{g_{AB}(0)}{\pi},
	\end{align}
	where $Cs\ll1$ and $\alpha^2 h(0)^2\gg 16\pi^2 C(2\alpha+1)$ due to the conditions of Eqs.~(\ref{eq:2cbcond1}) and (\ref{eq:2cbcond2}). Consequently, $h(s)$ behaves as
	\begin{align}
		h(s)^2\simeq\frac{16\pi^2\frac{2\alpha+1}{\alpha^2 }}
			{s+\frac{16\pi^2(2\alpha+1)}{\alpha^2h(0)^2}}.
	\end{align}
	We immediately notice that the solution is equal to Eq.~(\ref{eq:2bchrgsol}), which is the solution of $h(s)$ with vanishing two-body interaction. Therefore, $h(s)$ asymptotically behaves as $h(s)^2\simeq \frac{16\pi^2}{s}\frac{2\alpha+1}{\alpha^2}$ in the parameter region of Eq.~(\ref{eq:2cbcond3}). Following the discussion in Appendix~\ref{sec:app1} we finally arrive at the solution Eq.~(\ref{eq:2cbsolv4}) even in presence of the two-body interactions.\par
	As we noted in the main article, here the physical consequence of the two-body coupling constants are the infrared cutoffs of the semisuper-Efimov effect due to the condition Eq.~(\ref{eq:2cbcond1}).
	

\end{document}